\documentclass[a4paper,12pt]{article}
\usepackage[utf8]{inputenc} 
\usepackage[T1]{fontenc}
\usepackage[english]{babel} 
\usepackage{amsmath,amssymb,amsthm,mathrsfs,dsfont,mathtools,euscript} 
\usepackage{xcolor} 
\usepackage{authblk}  
\usepackage{ushort} 
\usepackage{url}
\usepackage{enumerate}
\usepackage[shortlabels]{enumitem} 
\usepackage{scalerel} 
\usepackage{relsize}
\usepackage[top=3cm, bottom=2.5cm, left=2cm, right=2cm]{geometry}
\usepackage{graphicx}
\usepackage[colorlinks=true]{hyperref}
\hypersetup{urlcolor=blue, linkcolor=red, citecolor=blue}

\newcommand{\eP}{\mathbb{P}}
\newcommand{\iL}{\mathcal{L}}
\newcommand{\iE}{\mathcal{E}}
\newcommand{\E}{\mathbb{E}}
\newtheorem{theorem}{Theorem}[section]

\title{Bacterial Metabolic Heterogeneity: from Stochastic to Deterministic Models }

\author[1,3]{C. Graham
\thanks{Email: \texttt{carl.graham@polytechnique.edu}}}
\author[2]{J. Harmand\thanks{Email: \texttt{jerome.harmand@inrae.fr}}}
\author[1,4]{S. M\'el\'eard
\thanks{Email: \texttt{sylvie.meleard@polytechnique.edu}}}

\author[1]{J. Tchouanti
\thanks{Email: \texttt{josue.tchouanti-fotso@polytechnique.edu}}}

\affil[1]{CMAP, CNRS, \'Ecole Polytechnique, IP Paris, 91128 Palaiseau, France}
\affil[2]{LBE-INRAE, 102 Avenue des \'Etangs, 11100 Narbonne, France}
\affil[3]{Inria}
\affil[4]{Institut Universitaire de France}

\date{May 15, 2020}

\begin{document}

    \maketitle
    
    \begin{abstract}
        We revisit the modeling of the diauxic growth of a pure micro\-organism on two distinct sugars which was first described by Monod. Most available models are deterministic and make the assumption that all cells of the microbial ecosystem behave homogeneously with respect to both sugars, all consuming the first one and then switching to the second when the first is exhausted. We propose here a stochastic model which describes what is called ``metabolic heterogeneity''. It allows to consider small populations as in microfluidics as well as large populations where billions of individuals coexist in the medium in a batch or chemostat. We highlight the link between the stochastic model and the deterministic behavior in real large cultures using a large population approximation. Then the influence of model parameter values on model dynamics is studied, notably with respect to the lag-phase observed in real systems depending on the sugars on which the microorganism grows. It is shown that both metabolic parameters as well as initial conditions play a crucial role on system dynamics.
    \end{abstract}

    \section{Introduction}
        Described for the first time by Monod~\cite{monod}, the diauxic growth consists in a biphasic growth in a bacterial population consuming two different sugars in a closed medium. The corresponding curve of biomass density at the macroscopic scale shows two distinct exponential phases separated by a ``plateau'' called lag-phase. The explanation proposed by Monod is that the preferred sugar (which is in some sense ``easier'' to metabolize) is consumed first while the metabolic pathway allowing the consumption of the second one is suppressed. When the concentration of the first sugar becomes low enough, this repression is lifted. Then, the microorganism may produce the enzymes necessary to metabolize the second sugar: this is the lag-phase. The second exponential growth is observed until the second sugar is eventually consumed.

        \begin{figure}[h]
            \begin{center}
                \includegraphics[width=10cm, height=7cm]{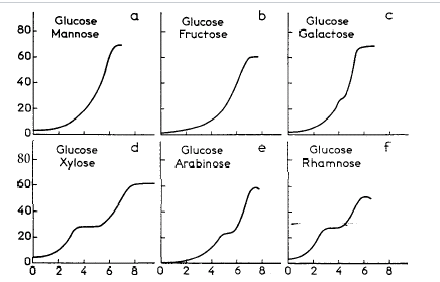}
                \caption{Growth of \emph{Escherichia coli} in the presence of different carbohydrate pairs serving as the only source of carbon in a synthetic medium (Monod~\cite{monod})}
                \label{fig-monod-orig-curves}
            \end{center}
        \end{figure}
        Until recently, it was admitted that the explanation given above was homogeneous within the cell population in the sense that each individual adopted exactly the same behavior at the same time: each cell first consumed the sugar that was ``easiest to metabolize''  first, then the other one after a duration corresponding to the lag-phase. Such an assertion implies that the latency time would simply be a constant depending only on the sugars involved. 
        In order to better understand this phenomenon and test hypotheses, many models of diauxic growth have been proposed in the literature (\cite{model1, model2, model3}). All such models have in common to make the hypothesis that each cell of the microorganism under consideration exhibits the same behavior with respect to both substrates at a given time. In addition, most approaches make use of deterministic models that are not suited for low biomass densities. 
        
        However, recent investigations suggest that lag phases are controlled by the inoculum history and organized with heterogeneity among individual cells (Bertrand~\cite{bertrand}). This fact was called ``metabolic heterogeneity''. Takhaveev and Heinemann~\cite{vakHein} suggested that this heterogeneity could be induced by mechanisms linked to ecological factors, gene expression, and other inherent dynamics, or by interaction between individuals, which all also depend on environment changes. 
        
        In this paper, following the idea that there is an intrinsic heterogeneity of cells within the ecosystem -- which yields a metabolic heterogeneity -- we develop a stochastic model of diauxic growth. More precisely, we propose a model of a batch culture of a pure strain growing on two different sugars. In this model, the metabolic heterogeneity is modeled via the possible emergence of a subpopulation able to consume the second sugar while the first one is not yet totally consumed. In other words, all cells do not exhibit the same behavior with respect to each substrate at a given time. To be as close as possible to the observations, the model accounts for the fact that in such situations the acetate produced -- which is a growth-inhibiting metabolite -- is co-consumed by each cell, as shown by 
        Enjalbert et al.~\cite{brice2017}. One goal of this work is to link both stochastic and deterministic approaches in order to explain the observations available at different scales, and to study the main parameters that control the length of the lag-phase.

        The paper is organized as follows.
        First, the stochastic model is presented. 
        Secondly, its behavior for large populations is approximated, allowing us to write a model consisting in a set of deterministic differential equations. 
        Then, the model is used to investigate the role of a number of model parameters and of initial conditions on the substrate consumption dynamics and on the length of the lag-phases.
        Eventually the main conclusions and perspectives are drawn.
        An appendix provides some additional information on proofs and simulations.
    
    \section{The stochastic model}
        First and foremost, let us introduce the parameter $K>0$ that scales the initial number of individuals. Indeed, the population size varies widely between different kinds of bacterial cultures, and may range from a few individuals in microfluidics (then $K$ is very small) to billions or more in fermenters (then $K$ is very large). 
        
        Let us consider two different substrates, Sugar1 which is preferential and Sugar2, and a stochastic population of bacteria split into two compartments constituted respectively of Sugar1 consumers and of Sugar2 consumers. 
        Let $N^K_1(t)$  and $N^K_2(t)$ denote respectively the numbers of individuals in each compartment and
        \[
            N^K(t)=\bigl(N^K_1(t),N^K_2(t)\bigr)\,, \quad t\ge0\,.
        \]    
        Here we are introducing the specific scaling in which $K$ can be seen as proportional to the carrying capacity of the medium and $1/K$ as proportional to the individual biomass, and in order to capture the two subpopulation densities we  introduce the rescaling
        \begin{equation*}
          n^K(t) = \bigl( n^K_1(t),n^K_2(t) \bigr) = \left( \frac{N^K_1(t)}{K}, \frac{N^K_2(t)}{K} \right), \quad t\ge0\,.
        \end{equation*}
        The mass concentration of each sugar is described by a continuous process
        \begin{equation*}
            R^K(t) = \bigl( R^K_1(t),R^K_2(t)\bigr)\,, \quad t\ge0\,,
        \end{equation*}
        which corresponds in the same order to Sugar1 and Sugar2. We also take into account the mass concentration $A^K(t)$ of a metabolite produced during the consumption of sugars by each individual and co-consumed with them. As an illustration, we may consider a mixed medium with glucose and xylose as Sugar1 and Sugar2, and acetate as the metabolite. 
       
        We describe the complete culture medium by the Markov process
        \begin{equation}
        \label{Markov-process}
            (n^K(t),R^K(t),A^K(t))_{t\ge0} = (n^K_1(t),n^K_2(t),R^K_1(t),R^K_2(t),A^K(t))_{t\ge0}
        \end{equation}
        evolving as follows.
        \begin{itemize}
            \item \textbf{Demography}. An individual growing on Sugar1 divides at rate $b_1(R^K_1(t),A^K(t))$ due to Sugar1 and metabolite co-consumption. Likewise, an individual growing on Sugar2 divides at rate $b_2(R^K_2(t),A^K(t))$ due to Sugar2 and metabolite co-consumption. This results in the jumps 
            \[
            \begin{aligned}
            n_1&\longrightarrow n_1 + \frac{1}{K} &&\text{at rate} &&  b_1(R^K_1(t),A^K(t))n_1\,, 
            \\
            n_2 &\longrightarrow n_2 + \frac{1}{K} &&\text{at rate} &&  b_2(R^K_2(t),A^K(t))n_2\,.
            \end{aligned}
            \]

            \item \textbf{State transitions}. An individual growing on Sugar1 switches its state in order to consume Sugar2 at rate $\eta_1(R^K(t))$,
            which depends on both resources since this is inhibited by the catabolic repression due to Sugar1, the preferential sugar. Likewise, an individual growing on Sugar2 switches its metabolic state to consume Sugar1 at rate $\eta_2(R^K_1(t))$ which depends only on the abundance of Sugar1. This results in the jumps 
            \[
            \begin{aligned}
                (n_1,n_2) &\longrightarrow \left(n_1 - \frac{1}{K},n_2+\frac{1}{K}\right) && \text{at rate} && \eta_1(R^K(t))n_1\,,
                \\
                (n_1,n_2) &\longrightarrow \left(n_1 + \frac{1}{K},n_2-\frac{1}{K}\right) && \text{at rate} &&\eta_2(R^K_1(t))n_2\,.
            \end{aligned}
            \]

            \item \textbf{Resource dynamics}. These are linked to the biomass and metabolite synthesis by the biochemical reactions that happen inside the batch. An individual growing on Sugar1 consumes continuously a small amount $\mu_1(R^K_1(t),A^K(t))/q_1K$ of Sugar1 and produces continuously a small amount $\theta_1\mu_1(R^K_1(t),A^K(t))/K$ of metabolite per unit of time, before dividing as a result of  this consumption. Similarly, an individual growing on Sugar2 consumes continuously a small amount $\mu_2(R^K_2(t),A^K(t))/q_2K$ of Sugar2 and produces continuously a small amount $\theta_2\mu_2(R^K_2(t),A^K(t))/K$ of metabolite per unit of time, before dividing as a result of  this consumption. Finally, each individual consumes a small amount $\mu_3(A^K(t))/q_3K$ of metabolite per unit of time before dividing as a result of  this consumption.  This leads us to describe the resource dynamics by the dynamical system         
            \begin{equation}
            \label{equadiff-resource-sto}
            \left\{
            \begin{aligned}
                \frac{dR^K_1}{dt}(t) &= - \frac{\mu_1(R^K_1(t),A^K(t))}{q_1 V}n^K_1(t)\,, 
                \\
                \frac{dR^K_2}{dt}(t) & = - \frac{\mu_2(R^K_2(t),A^K(t))}{q_2 V}n^K_2(t)\,, 
                \\
                \frac{dA^K}{dt}(t) &= - \frac{\mu_3(A^K(t))}{q_3 V}\bigl(n^K_1(t)+n^K_2(t)\bigr) + \frac{\theta_1}{V} \mu_1(R^K_1(t),A^K(t))n^K_1(t) 
                \\
                &\qquad+ \frac{\theta_2}{V} \mu_2(R^K_2(t),A^K(t))n^K_2(t)\,.
            \end{aligned}
            \right.
            \end{equation}
        \end{itemize} 
        The typical situation we will consider is the following. The initial conditions satisfy
        \begin{displaymath}
            n^K(0) = \left( \frac{\lfloor n^0_1K\rfloor}{K}, \frac{\lfloor n^0_2K\rfloor}{K} \right), \quad R^K(0) = r^0\,, \quad A^K(0) = 0\,,
        \end{displaymath}
        in which $(n^0,r^0) = (n^0_1,n^0_2,r^0_1,r^0_2)$ is fixed. The above rate functions involve Monod-type  and classic inhibition functions and are of the forms
        \begin{align*}
            \mu_j(r_j,a) &= \bar{\mu}_j\frac{r_j}{\kappa_j + r_j}\cdot\frac{\lambda}{\lambda + a}\,, \quad j=1,2\,, 
            &\mu_3(a) &= \bar{\mu}_3\frac{a}{\kappa_3 + a}\cdot\frac{\lambda}{\lambda + a} \,,
            \\
            \eta_1(r) &= \bar{\eta}_1\frac{r_2}{k_1 + r_2}\cdot\frac{k_i}{k_i + r_1}\,, \qquad 
            &\eta_2(r_1) &= \bar{\eta}_2\frac{r_1}{k_2 + r_1}\,.
        \end{align*}
        In addition, the choice
        \begin{equation}\label{div}
            b_j(r_j,a) = \mu_j(r_j,a) + \mu_3(a)\,, \quad j=1,2\,,
        \end{equation}
        ensures a conservation law on average:
        \begin{equation}
            \E\left\{ n^K_1(t) + n^K_2(t) + V\left[ q_1(1+\theta_1q_3)R^K_1(t) + q_2(1+\theta_2q_3)R^K_2(t) + q_3A^K(t)\right] \right\} = C^{\mathrm{st}}.
        \end{equation}
        To illustrate this model, we use the  parameters described in Table~\ref{tab:para-rate-fnct} taken from recent batch experiments (Barthe et al.~\cite{Barthe2020}) for the sugars glucose and xylose.
        \begin{table}[h!]
        \centering
            \begin{tabular}{|clc|}
                \hline
                Parameter & Biological signification & Default value \\
                & & for simulations \\
                \hline
                $\bar{\mu}_1$ & Maximal growth rate on Sugar1 & 6.50e-01  \\
                $\kappa_1$ & Monod constant on Sugar1 & 3.26e-01 \\
                $\lambda$ & Inhibition coefficient due to the metabolite & 4.70e-01 \\
                $\bar{\mu}_2$ & Maximal growth rate on Sugar2 & 5.70e-01  \\
                $\kappa_2$ & Monod constant on Sugar2 & 4.68e-01  \\
                $\bar{\mu}_3$ & Maximal growth rate on the metabolite & 1.47e-01  \\
                $\kappa_3$ & Monod constant on the metabolite & 6.45e-01  \\
                $\bar{\eta}_1$ & Maximal switching rate from Sugar1 to Sugar2 & 2.04e-03  \\
                $k_1$ & Regulation coefficient of the Sugar1 to Sugar2 transition & 1.20e-02  \\
                $k_i$ & Inhibition coefficient of the Sugar1 to Sugar2 transition & 1.03e-03  \\
                $\bar{\eta}_2$ & Maximal switching rate from Sugar2 to Sugar1 & 6.60e-01  \\
                $k_2$ & Regulation coefficient of the Sugar2 to Sugar1 transition & 4.50e-02  \\
                $q_1$ & Individual yield on Sugar1 & 5.50e-01 \\
                $\theta_1$ & Metabolite yield on Sugar1 & 6.00e-01  \\
                $q_2$ & Individual yield on Sugar2 & 4.50e-01  \\
                $\theta_2$ & Metabolite yield on Sugar2 & 5.60e-01   \\
                $q_3$ & Individual yield on the metabolite & 2.50e-01   \\
                $V$ & Bioreactor volume & 1.0  \\
                $n^0$ & Initial subpopulation densities & (2.80e-01, 0.0) \\
                $r^0$ & Initial sugar concentration & (8.15, 9.05) \\
                \hline 
            \end{tabular}
            \caption{Parameters for the rate functions in the example.}
            \label{tab:para-rate-fnct}
        \end{table}
        
\begin{figure}[h!]
            \begin{center}
                \includegraphics[width=7cm, height=5cm]{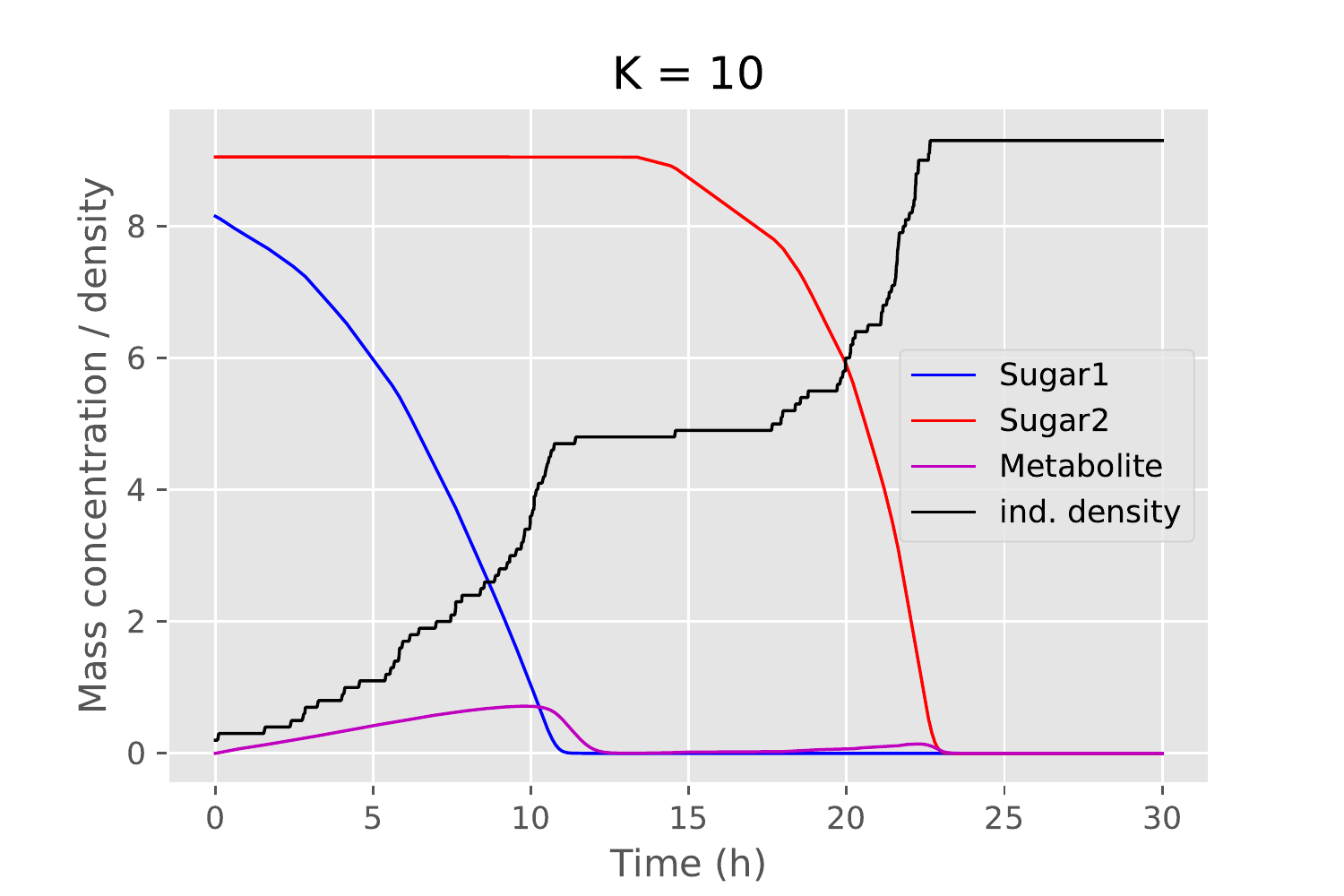}
                \includegraphics[width=7cm, height=5cm]{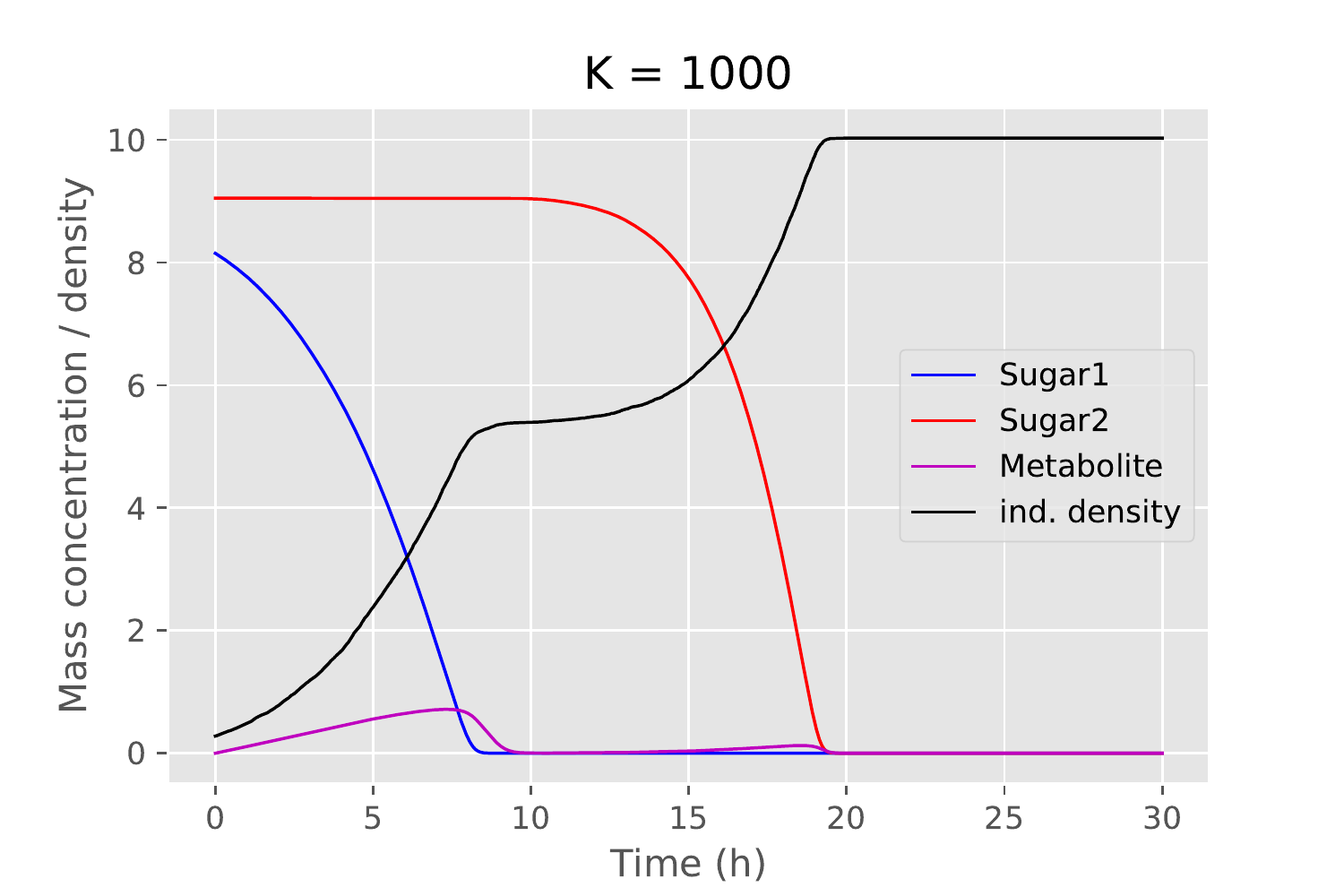}
                \caption{Total population densities and resource concentrations for a small ($K=10$) and a moderately large ($K=1000$) population.}
                \label{fig-tot-pop-dens}
            \end{center}
        \end{figure}

        Figure~\ref{fig-tot-pop-dens} shows that this model is able to predict the diauxic growth observed by Monod (see Figure~\ref{fig-monod-orig-curves}). This can be observed  even for a small number of individuals. 
        We additionally observe that the trajectories oscillate randomly for small $K$ and become smoother as K becomes larger. This observation will be developed in the next section, in which the stochastic model  will be shown to be approximated by a deterministic model when $K$ increases to infinity.
 %       
        %\bigskip
  %      
        
    \section{Large population approximation}
        The amplitude of any jump occurring in the population is bounded by a factor of  the weight $1/K$ attributed to a single individual and hence has a variance of order $1/K^2$. Moreover, the mean number of jumps per unit of time is of order $K$, the order of magnitude of the number of individuals. Heuristically, for a large population the process should approach a limit deterministic continuous process dictated by the mean  values, with random oscillations around this limit corresponding to variances of order $1/K$ and hence to standard deviations  of order $1/\sqrt{K}$. We can build on these heuristics and prove that  the stochastic model is indeed approximated by a deterministic model in the limit of large $K$. This yields the following deterministic limit. 
        \begin{theorem}
        \label{thm-convergence}
            Let us assume that 
        	\[
                \forall \varepsilon>0\,, \quad \eP\Bigl( \bigl\lVert \bigl(n^K(0),R^K(0),A^K(0)\bigr) - \bigl(n^0,r^0,a^0\bigr) \bigr\rVert > \varepsilon \Bigr)\xrightarrow[K\to\infty]{}0\,,
            \]
            and that $\sup_{K}\mathbb{E}\Bigl(\bigl\lVert \bigl(n^K(0), R^K(0), A^K(0)\bigr) \bigr\rVert \Bigr)<+\infty$.
            Let $(n(t),r(t),a(t))_{t\geq 0}$ be the unique solution with initial condition $(n^0,r^0,a^0)$ of the differential system
            \begin{equation}
            \label{syst}
            \left\{
            \begin{aligned}
                n'_1(t) &= \big\{ b_1(r_1(t),a(t)) - \eta_1(r(t)) \big\}n_1(t) + \eta_2(r_1(t))n_2(t)\,, 
                \\
                n'_2(t) &= \big\{ b_2(r_2(t),a(t)) - \eta_2(r_1(t)) \big\}n_2(t) + \eta_1(r(t))n_1(t) \,,
                \\
                r'_1(t) &= - \frac{\mu_1(r_1(t),a(t))}{q_1V}n_1(t)\,, 
                \\
                r'_2(t) &= - \frac{\mu_2(r_2(t),a(t))}{q_2V}n_2(t)\,, 
                \\		
                a'(t) &= - \frac{\mu_3(a(t))}{q_3V }(n_1(t)+n_2(t)) + \frac{\theta_1 \mu_1(r_1(t),a(t))n_1(t) + \theta_2 \mu_2(r_2(t),a(t))n_2(t)}{V} \,.
            \end{aligned}
            \right.
            \end{equation}
            Then the stochastic process $(n^K(t),R^K(t),A^K(t))_{t\geq 0}$ is approximated by $(n(t),r(t),a(t))_{t\geq 0}$
            for large $K$ in the sense that
            \begin{equation*}
                \forall T>0, \varepsilon>0\,,
                \quad
                \eP\Biggl(\sup_{0\leq t\leq T} \bigl\lVert (n^K(t),R^K(t),A^K(t)) - (n(t),r(t),a(t)) \bigr\rVert > \varepsilon 
                \Biggr)
                \xrightarrow[K\to\infty]{}0\,.
            \end{equation*}
        \end{theorem}
        This theorem allows to explain on a rigorous basis the observations we have made on the simulations in the previous section.
        Before discussing the proof methods, let us address the important question of the range of validity of the approximation.
        
        For small $K$ and most notably for populations consisting of a few individuals, the deterministic system is not a good approximation of the stochastic model and does not provide a pertinent model for the population. On the contrary, when $K$ is large enough for  the approximation to be accurate, the deterministic system provides a pertinent model on which  theoretical studies and numerical computations can be performed for qualitative and quantitative investigations on the population.
        
        Therefore, it is fundamental to obtain a precise evaluation of the size of $K$ required for the approximation to be tight and to assess the error made in terms of $K$. The heuristics given before the theorem indicate that that the error terms should be of order $1/\sqrt{K}$. Under adequate assumptions on the initial conditions, this can be made rigorous through a functional central limit theorem: the process
        \[
            \sqrt{K}\Bigl( \bigl(n^K(t),R^K(t),A^K(t)\bigr) - (n(t),r(t),a(t)) \Bigr)\,, \quad t\ge0\,,
        \]
        converges as $K$ goes to infinity to a Gaussian process of Ornstein-Uhlenbeck type, with mean and covariance structure expressed solely in terms of the limit process $(n(t),r(t),a(t))_{t\ge0}$ and of the variance of the jumps in a sufficiently explicit fashion to be well evaluated.
        This allows to evaluate the minimal size of $K$ required for a tight approximation and to provide confidence intervals on this, as well as the possibility for intermediate sizes of $K$ to simulate the deterministic limit process and add to it fluctuations simulated according to this Gaussian process in order to obtain a tighter approximation.
        
        The proofs of  Theorem~\ref{thm-convergence} and of the functional central limit theorem build on the heuristic explanation given before the theorem using probabilistic compactness-uniqueness methods. 
        Ethier and Kurtz~\cite{etk} is a classic book on the subject, and Anderson and Kurtz~\cite{Anderson2015} and 
        Bansaye and M\'el\'eard~\cite{sylVin} provide pedagogical expositions well suited to the present field of application.

\begin{figure}[h!]
            \begin{center}
                \includegraphics[width=7cm, height=5.5cm]{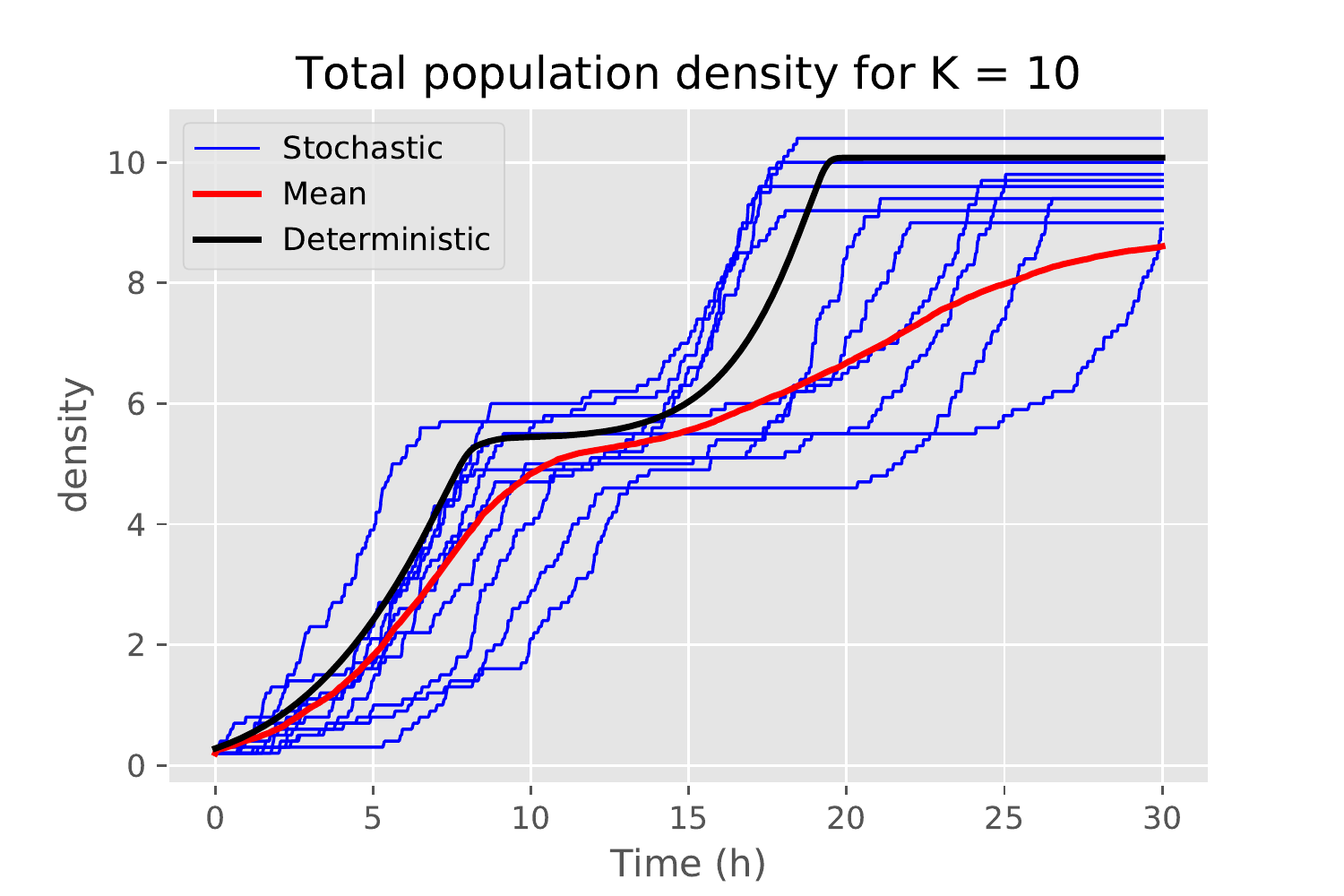}
                \includegraphics[width=7cm, height=5.5cm]{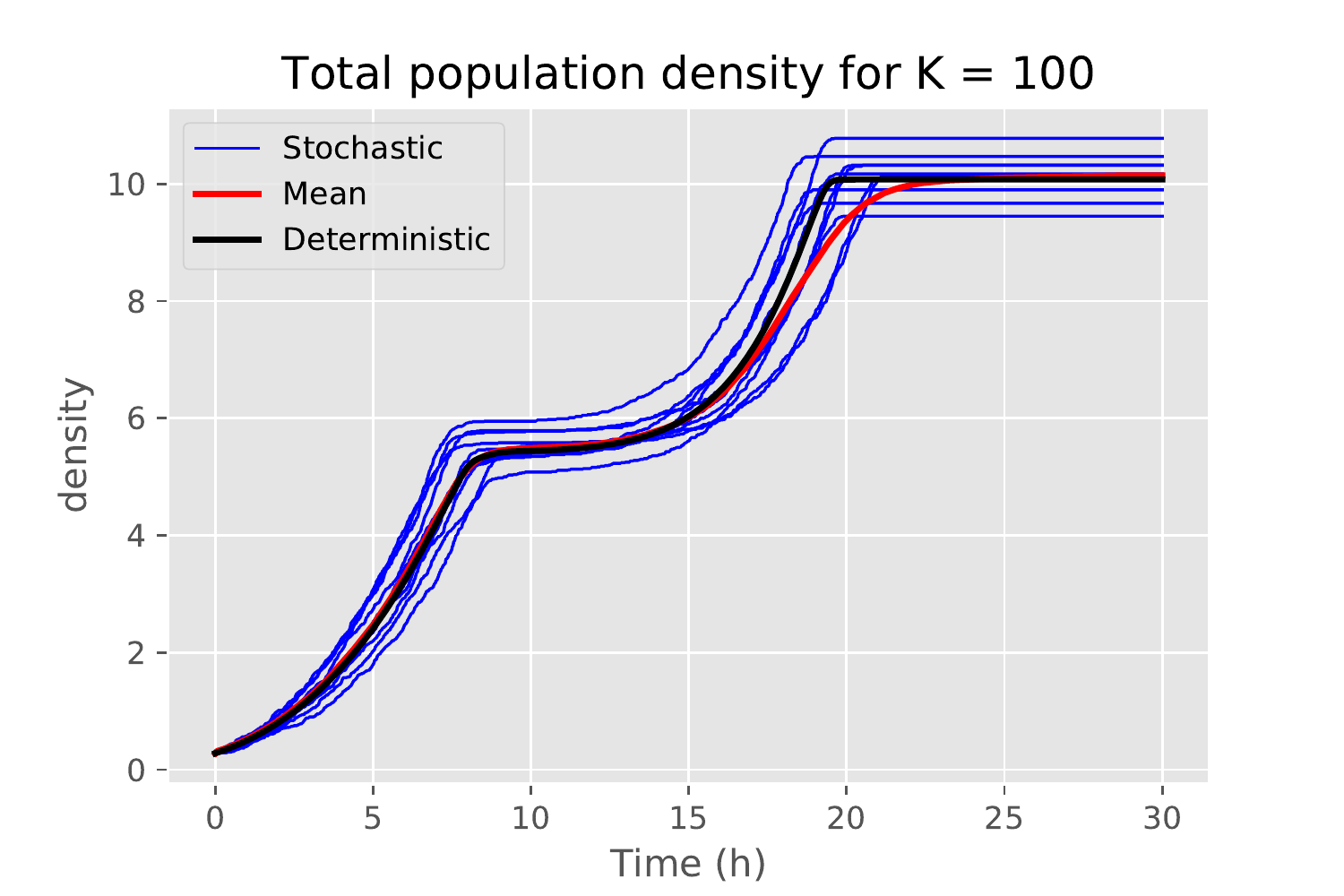}
                \includegraphics[width=7cm, height=5.5cm]{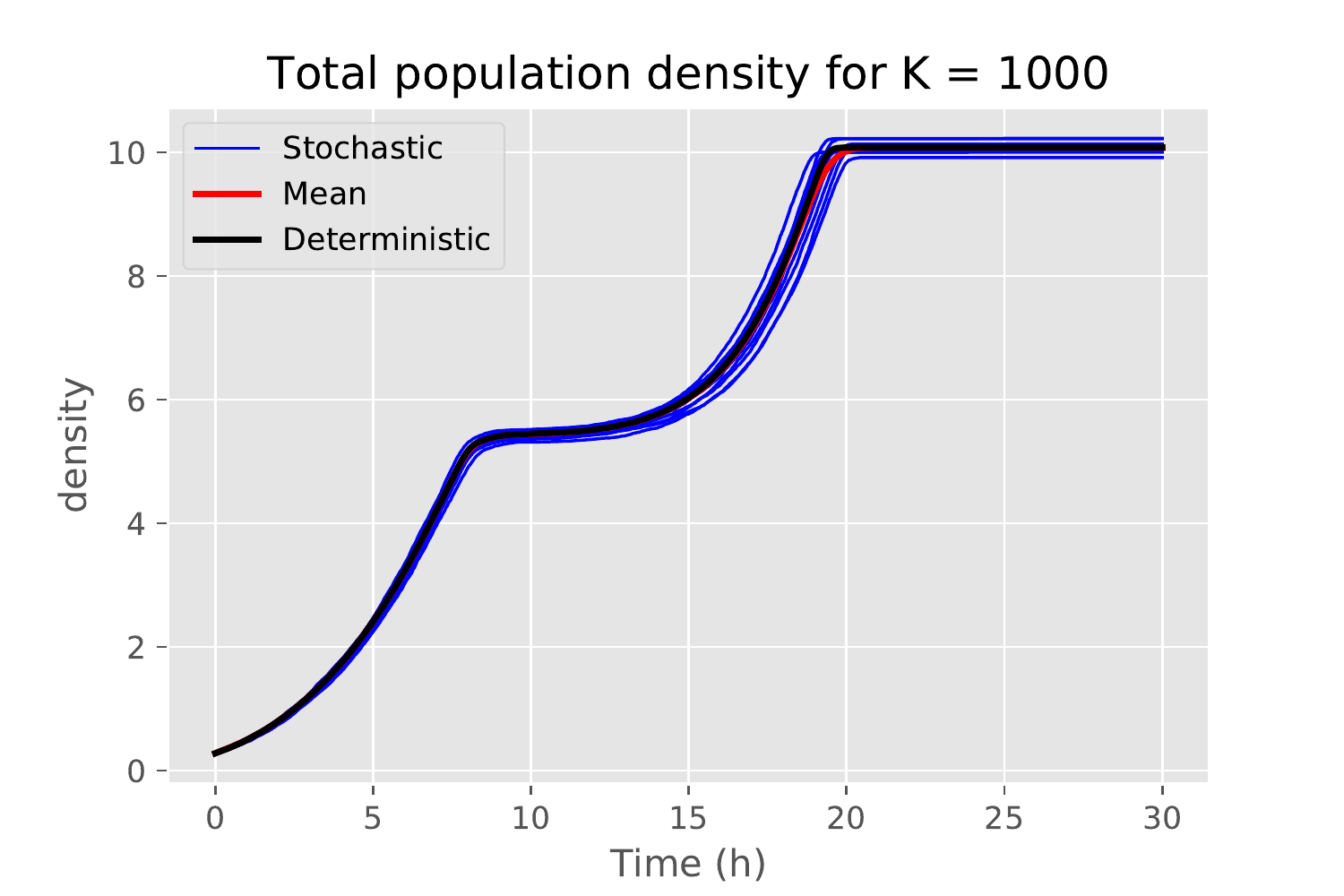}
                \caption{Ten independent stochastic trajectories, the empirical mean over a hundred independent trajectories, and the deterministic limit simulated for each of the total populations in the cases $K=10$, $K=100$, and $K=1000$.}
                \label{fig-100-sim-3-K}
            \end{center}
        \end{figure}

        We illustrate these convergence results in Figure~\ref{fig-100-sim-3-K}, by the simulations of a hundred trajectories of the total biomass for the stochastic and the limiting model, for three increasing values of the scale parameter $K$. 
        
    \section{Heterogeneity and lag-phase sensitivity}
\begin{figure}[h!]
            \begin{center}
                \includegraphics[width=17cm, height=6cm]{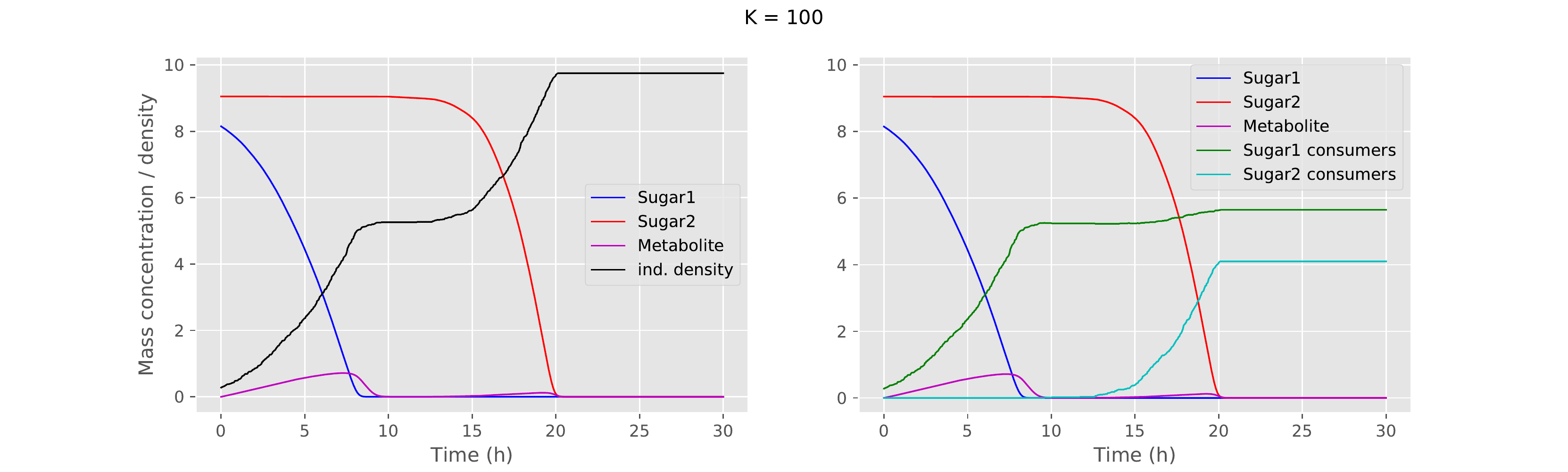}
                \includegraphics[width=17cm, height=6cm]{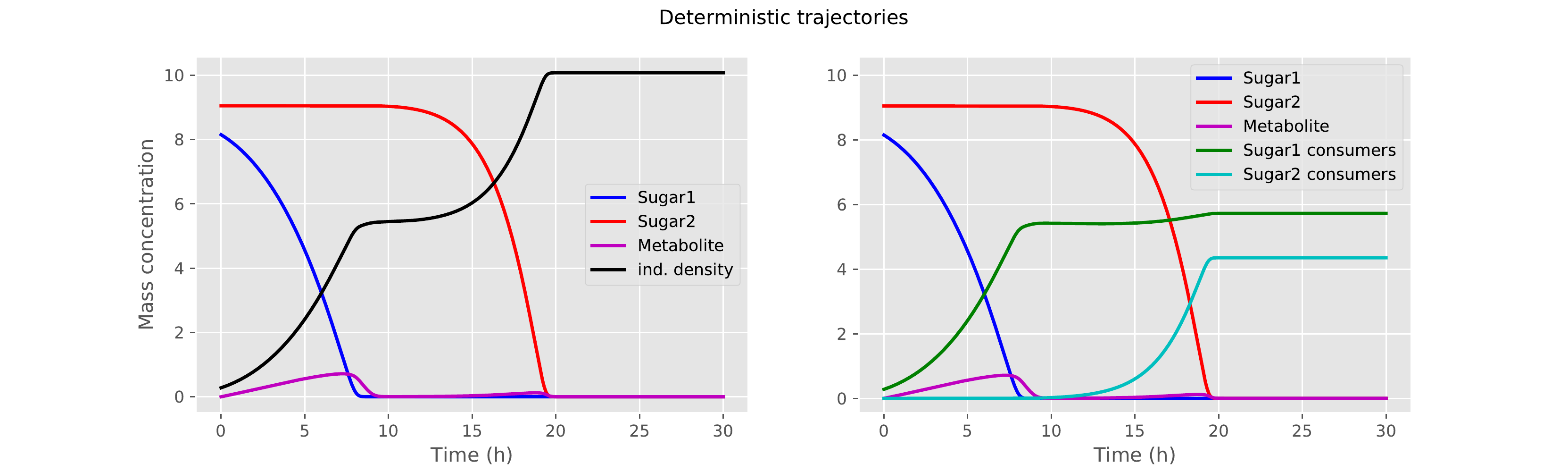}
                \caption{Metabolic Heterogeneity. Diauxic growth of the total population (left) and growths of the subpopulations (right)  for a simulation of the stochastic model with $K=100$ (top) and of the deterministic limit (bottom). The evolutions of the two resources are plotted for each.}
                \label{fig-Det-subpop}
            \end{center}
        \end{figure}

        As shown in Figure \ref{fig-Det-subpop}, the model is able to capture the heterogeneity of the population observed by biologists as well as the diauxic growth at the level of the  total population size highlighted by Monod.
        
        We present several simulations, each with either one varying metabolic parameter or a varying initial condition. These varying factors have been carefully chosen after a numerical exploration showing that they have a strong influence on the lag-phase duration, and in particular that they allow to reproduce dynamics similar to the ones obtained by Monod, see Figure~\ref{fig-monod-orig-curves}.
        
        \subsection{Influence of the metabolic parameters on the diauxie length}
\begin{figure}[h!]
                \begin{center}
                    \includegraphics[width=8cm, height=6cm]{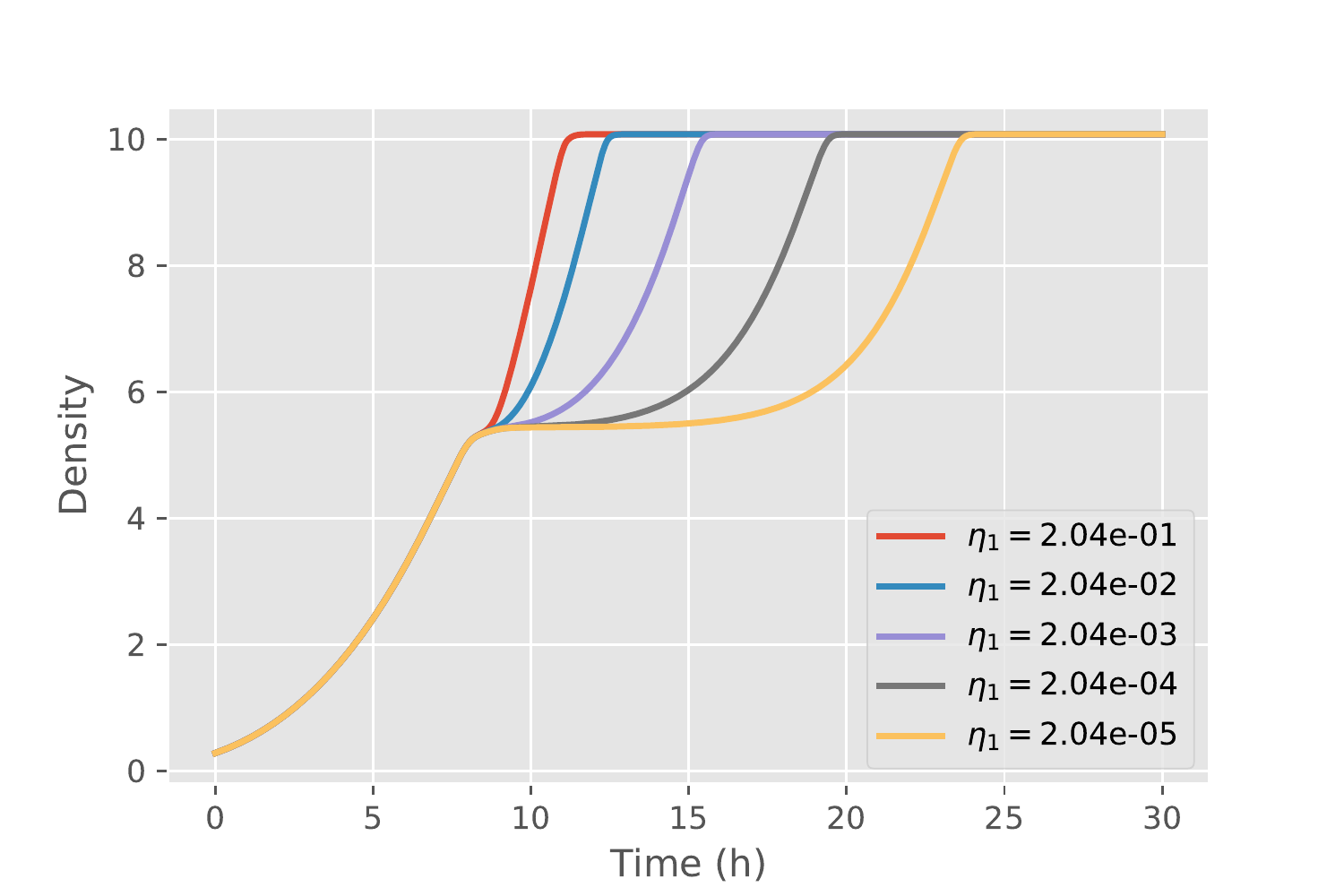}
                    \includegraphics[width=8cm, height=6cm]{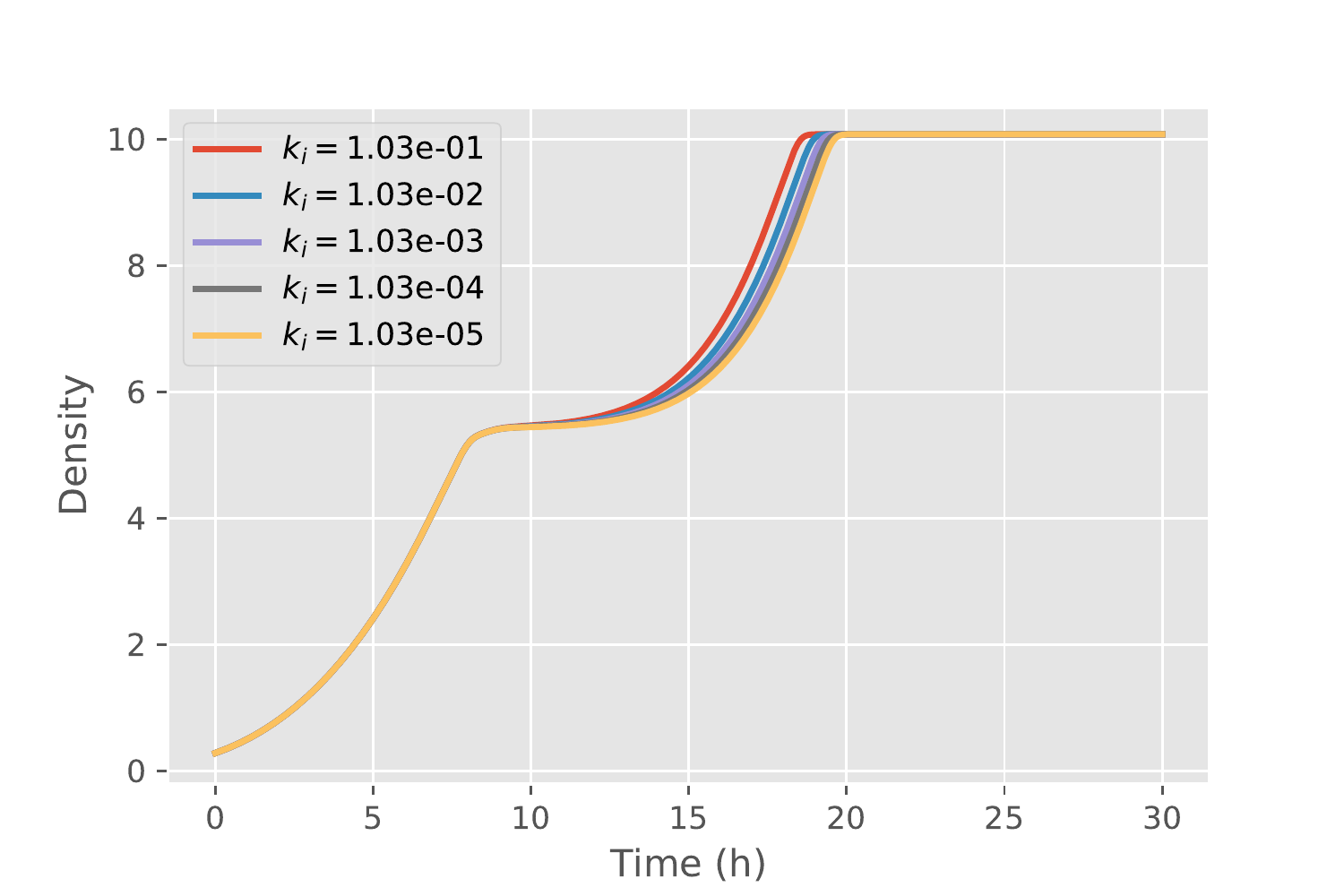}
                    \caption{Effects of changes in the switch parameter $\bar{\eta}_1$ (left) and in the inhibition coefficient $k_i$ (right) on the diauxic growth.}
                    \label{fig-Det-Eta1}
                \end{center}
            \end{figure}

            We here consider two major parameters:
            \begin{itemize}
                \item The maximal switching rate $\bar{\eta}_1$ from Sugar1 consumption to Sugar2 consumption, 
                which describes individual switches on high Sugar2 medium after Sugar1 exhaustion.  
                As this parameter increases, the transitions become more frequent and the lag phase shorter.
                
                \item The inhibition coefficient $k_i$ of the Sugar1 to Sugar2 transition, which describes the catabolic repression due to the abundance of the preferential sugar. Due to its role in the inhibition functions, as this parameter decreases the transitions become more frequent and the lag phase shorter.
            \end{itemize}
            Our simulations are represented in Figure \ref{fig-Det-Eta1}.

        \subsection{Influence of the initial conditions on the diauxie length}        
\begin{figure}[h!]
                \begin{center}
                    \includegraphics[width=8cm, height=6cm]{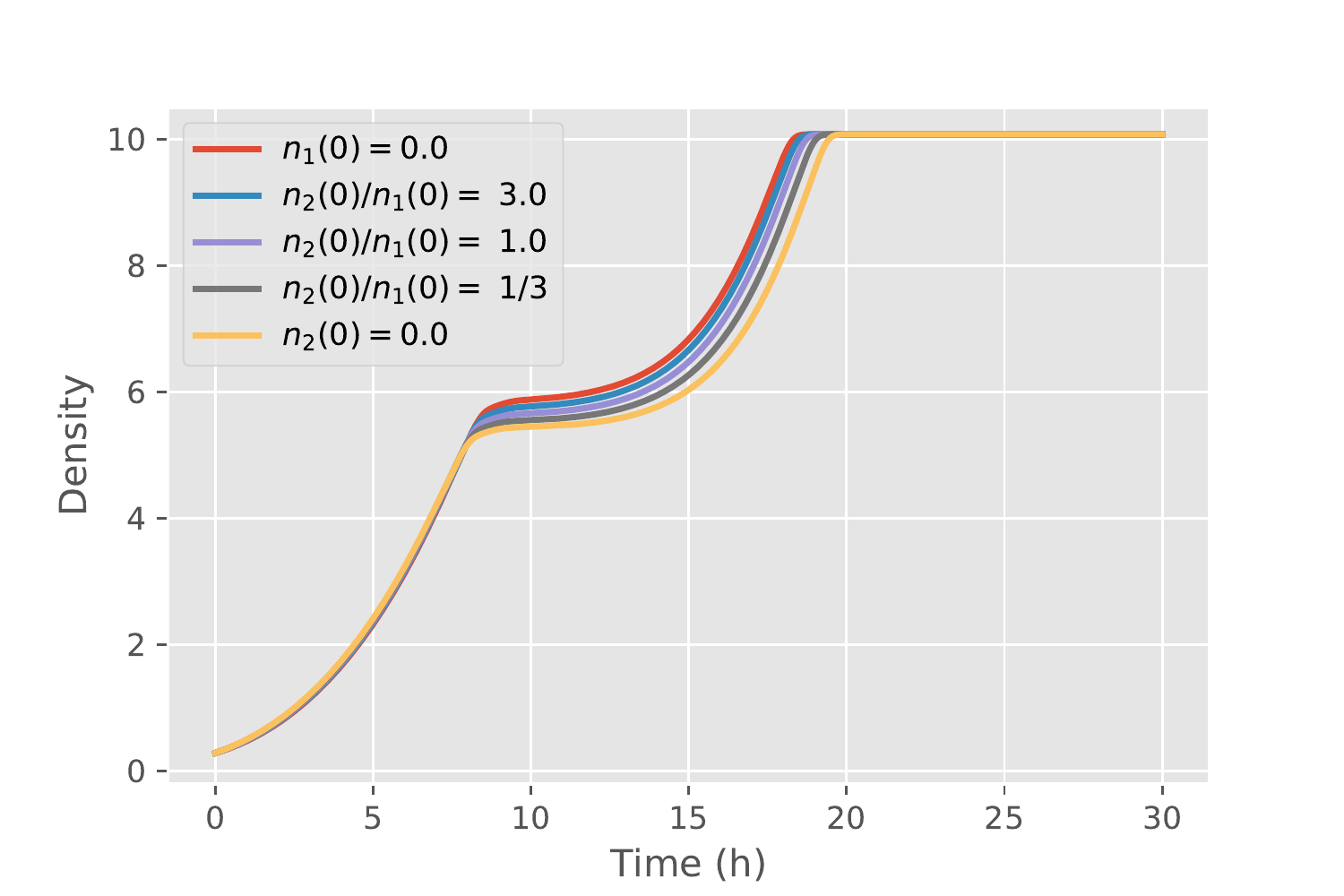}
                    \caption{Effects of changes in the initial condition on the diauxic growth.}
                    \label{fig-Det-Init}
                \end{center}
            \end{figure}

            The initial condition $n(0)=(n_1(0),n_2(0))$ of the population also has a strong influence on the duration of the lag-phase. Indeed, if the population density of Sugar2 consumers is significant with respect to the density of Sugar1 consumers in the initial condition, then the lag-phase will be short.  Our simulations are represented in Figure \ref{fig-Det-Init}.
            
    \section{Conclusion}
        In this paper, we proposed a stochastic model of diauxic growth of a microorganism on two different sugars. The model assumes that  the individuals preferentially consume  one of the sugars while the metabolic pathway allowing the consumption of the second one is repressed until the first sugar is exhausted. To account for the fact that all individual do not behave homogeneously with respect to the consumption of sugars - which is called metabolic heterogeneity - it is supposed that some individuals can switch their metabolism in such a way they can consume the second sugar while the first one is not totally exhausted. Thus the model involves two different subpopulations: the first one which grows on the first sugar, and the second one, which emerges from the first subpopulation and consumes the second sugar. In addition, three resource variables with continuous dynamics are added: the two sugars, and the intermediate metabolite which is produced when the sugars are consumed and then re-consumed by both subpopulations. Then, the deterministic model that approximates the stochastic model dynamics is derived using a large population approximation. Using parameter values that are supposed to be close to those we can find in real experiments, for instance when \emph{E.~coli} grows on both glucose (the preferential sugar) and xylose, we performed a number of simulations in order to investigate the influence of the most important parameters on the model dynamics. Further, we show the importance of the weighting factor $K$, which allows us to understand what is the population size starting from which the deterministic model can be used to approximate the stochastic model dynamics. Finally, it is shown that several parameters, such as the maximal switching rate $\bar{\eta}_1$ from Sugar1 to Sugar2 consumption and the inhibition coefficient $k_i$ of the Sugar1 to Sugar2 transition, as well as the initial conditions of the system significantly influence the lag-phase, allowing us to pave the way and to suggest strategies to minimize the lag-phase in practical experiments.
        
    \section*{Acknowledgments}
        We warmly thank our biologist colleagues Manon Barthe, Muriel Cocaign and Brice Enjalbert who have brought this  problem to our attention and  shared fruitful discussions with us. This work has been supported by the Chair ``Mod\'elisation Math\'ematique et Biodiversit\'e'' of Veolia Environnement-\'Ecole Polytechnique-Mus\'eum national d'Histoire naturelle-Fondation X and by the ANR project JANUS (ANR-19-CE43-0004-01).
        
    \newpage    

%%%

%%%
    
    \appendix
    \section{Appendix}
    \subsection{Proof of main results}
        Let us comment the proof  of Theorem \ref{thm-convergence} which can easily adapted from the results in 
        \cite{etk, Anderson2015,sylVin}. 
        Let us firstly note that we can express the stochastic process $(n^K(t),R^K(t))_{t\geq 0}$ as 
        \begin{eqnarray*}
            n^K_{1}(t) &=&  n^K_{1}(0) + M^K_{1}(t)  \\
            && \hskip 0.5cm + \int_{0}^t \bigg(\big\{ b_1(R^K_1(s),A^K(s)) - \eta_1(R^K(s)) \big\} n^K_{1}(s) + \eta_2(R^K_1(s)) n^K_{2}(s)\bigg) ds\,, \\
            n^K_{2}(t) &=&   n^K_{2}(0) + M^K_{2}(t) \\
            && \hskip 0.5cm  + \int_{0}^t \bigg(\big\{ b_2(R^K_2(s),A^K(s)) - \eta_2(R^K_1(s)) \big\} n^K_{2}(s)+ \eta_1(R^K(s)) n^K_{1}(s)\bigg) ds\,,
        \end{eqnarray*}
        where the processes $M^K_{1}$ and $M^K_{2}$ are square integrable martingales such that
        \begin{align}
        \label{variance}
            &\mathbb{E}( (M^K_{1}(t))^2) = {1\over  K} \int_{0}^t   \bigg(\big\{ b_1(R^K_1(s),A^K(s)) + \eta_1(R^K(s)) \big\} n^K_{1}(s) + \eta_2(R^K_1(s)) n^K_{2}(s)\bigg)\, ds\,,\nonumber\\
            & \mathbb{E}( (M^K_{2}(t))^2) =  {1\over  K} \int_{0}^t \bigg(\big\{ b_2(R^K_2(s),A^K(s)) + \eta_2(R^K_1(s)) \big\} n^K_{2}(s)+ \eta_1(R^K(s)) n^K_{1}(s)\bigg)\, ds\,,\nonumber 
            \\
            & \mathbb{E}( M^K_{1}(t) M^K_{2}(t)) = -  {1\over  K} \int_{0}^t  \bigg( \eta_1(R^K(s)) \, n^K_{1}(s) +\eta_2(R^K_1(s)) \, n^K_{2}(s)\bigg)\, ds\,.
        \end{align}
        The proof firstly consists in showing that the sequence of  laws of  the stochastic processes $(n^K(t),R^K(t),A^K(t), t\geq 0)_{K}$ is relatively compact. It is based on 2-moments estimates, uniform on finite time intervals and on $K$ and on a well known criterion of uniform tightness (\emph{cf.}~for example \cite{sylVin}). Then there exists at least one limiting probability measure (on the path space). Using the fact that the jump amplitudes are going to $0$ when $K$ tends to infinity, uniformly on finite time intervals, we deduce that these probability measures only charge continuous trajectories. Moreover, the moment estimates and (\ref{variance}) allow to prove that the martingale part converges in probability  to $0$ when $K$ tends to infinity. Therefore, it is easy to deduce that the limiting probability measures only charge the solutions of the dynamical system (\ref{syst}). The last step consists in proving the uniqueness of such a solution, which is due to a Cauchy-Lipschitz Theorem. 
          
	\subsection{Numerical simulations}
        In order to simulate the Markov process $(X^K(t))_{t\ge0} =(n^K(t),R^K(t),A^K(t))_{t\geq 0}$ defined in \eqref{Markov-process} for various sets of parameters, we propose an algorithm simulating numerically the differential system satisfied by the resources in between the jump instants, while the jump instants and the jump amplitudes are simulated directly in terms of the past. The ideas are based on first principles according to the Markov property.
        
        The jump structure of $(X^K(t))_{t\ge0}$   can be described locally at each state $x$ by the value $\alpha(x)\ge0$ of a jump rate function  
        $\alpha$ and if $\alpha (x)>0$ by a probability measure $\pi(x,dh)$ for drawing the amplitudes of the jumps. More precisely, there are overall $p\ge1$ possible non-null jump amplitudes $h_1, \cdots,h_p$, taken at each state $x = (n,r,a)$ at respective rates $\alpha_1(x)\ge0, \cdots,\alpha_p(x)\ge0$, and         
        \[
            \alpha (x) = \sum_{i=1}^p\alpha_i(x)\,,\qquad\pi(x,h_i) = \frac{\alpha_i(x)}{\alpha(x)}\,,\; \; i=1,\cdots,p\,.
        \]

        The strong Markov property yields interesting consequences  for the construction of the process.
        The future of the process after each jump is independent from its past given the new state. Thus, in order 
        to construct the process it is sufficient to be able to do so from time $0$ until the first jump instant, and then 
        iterate the procedure by considering each jump instant as a new time origin. Moreover, starting at time $0$ 
        the probability that  the process $(X^K(t))_{t\ge0}$ has not jumped yet at time $t>0$ 
        is given in terms of the rate function $\alpha$ by
        \[
            \exp\biggl(-\int_0^t \alpha(X^K(s))\, ds\biggr)\,.
        \]
        
        This allows to construct the first jump instant as follows. If the non-decreasing continuous 
        process $(\Lambda(t))_{t\ge0}$ and its left-continuous inverse $(\Lambda^{-1}(t))_{t\ge0}$ are defined by
        \begin{equation}
        \label{Lambda-and-inv}
        \Lambda(t) =\int_0^t \alpha(X^K(s))\, ds\,,
        \qquad
        \Lambda^{-1}(t) = \inf\{u\ge 0 : \Lambda(u) \ge t\}\,,
        \end{equation}
        and $D$ is an exponential random variable of parameter $1$, then 
        \[
        \eP\bigl(\Lambda^{-1}(D) >t\bigr) 
        =\eP\biggl(D>\int_0^t \alpha(X^K(s))\, ds\biggl) 
        = \exp\biggl(-\int_0^t \alpha(X^K(s))\, ds\biggr)\,.
        \]
        Hence, we can simulate the first jump instant $T_1$ of the process by taking $T_1 = \Lambda^{-1}(D)$ while 
        simultaneously constructing the process on  $[0,T_1)$. If $X^K(T_1-) =x$ then $X^K(T_1) =x+h$ for a jump amplitude 
        $h$ drawn according to $\pi(x,dh)$.  
        
        Using this construction directly for an actual simulation raises several issues.
        
        The first problem is that we must be able to simulate the process $(X^K(t))_{t\ge0}$ up to the first jump instant. 
        In the present situation this consists in simulating the components  $(R^K(t),A^K(t))_{t\geq 0}$ of the 
        Markov process~\eqref{Markov-process} by solving the differential system~\eqref{equadiff-resource-sto} in which the other components of~\eqref{Markov-process} remain constant between jumps. This cannot be done exactly but can be approximated numerically quickly and with precision.
         
        The second problem is that simultaneously to $(R^K(t),A^K(t))_{t\geq 0}$ we must be able to compute the integral $\Lambda(t)$ and its inverse $\Lambda^{-1}(t)$ defined in~\eqref{Lambda-and-inv}. This can be done numerically but is often costly in computer time and inefficient. This has a practical solution which we proceed to describe. We introduce a function $\tilde{\alpha}$ such that $\alpha \le \tilde{\alpha}$ and that the corresponding $\tilde{\Lambda}$ and that $\tilde{\Lambda}^{-1}$ defined similarly to~\eqref{Lambda-and-inv} are simpler to compute than $\Lambda$ and $\Lambda^{-1}$. We simulate the process $(X^K(t))_{t\ge0}$  by an acceptance-rejection method which proposes a jump from state $x$ at rate $\tilde{\alpha}(x)$ and accepts it with probability  $\alpha(x)/\tilde{\alpha}(x)$ and else rejects it. There are various  ways to justify that this construction is correct. One of these is to consider the rejection as the introduction of a jump of amplitude $0$ taken at the excessive rate $\tilde{\alpha}(x)- \alpha(x)$ (the process does not actually jump, and this is called a ``fictitious jump'') and reason as above. The simplest situation is when the dominating function $\tilde{\alpha}$ is a constant. Then the true jump instants of $(X^K(t))_{t\ge0}$ constitute a thinning of a Poisson process of constant intensity $\tilde{\alpha}$, which can be easily simulated, in which a  jump  instant of this Poisson process taken when $X^K(T_1-) =x$ is accepted with probability $\alpha(x)/\tilde{\alpha}(x)$.
        
        Let us  come back to our model and denote by $\iL(X^K(0))$ the distribution of the initial random vector $X^K(0)$, by $\iE\left( \lambda\right)$  the  exponential law with parameter  $\lambda>0$ and by $\mathcal{U}([0,1])$ the uniform law on $[0,1]$. If we moreover denote  by $(\phi(x,t-t_0))_{t\geq t_0}$ the flow of the process $(X^K(t))_{t\geq 0}$ from an initial condition $X^K(t_0) = x$ until the next jump time, the above description can be summarized in the following algorithm.
        \begin{center}
            \begin{tabular}{|ll|}
                \hline
                \quad & \qquad \qquad \qquad \qquad \qquad \qquad \quad \textbf{Algorithm} \\
                \hline
                \quad & \quad \\
                \quad & Simulate $x_0\thicksim \iL(X^K(0))$  \\
                \quad & $T_0 \longleftarrow 0$ ;  \\
                \quad & $k \longleftarrow 0$ ;  \\
                \quad & \textcolor{blue}{Repeat}  \\
                \quad & \qquad Simulate $\epsilon_{k+1}\thicksim\iE\left( \tilde{\alpha}\left(x_k \right)\right)$ ;  \\
                \quad & \qquad $T_{k+1} \longleftarrow T_{k} + \epsilon_{k+1}$ ; \\
                \quad & \qquad Follow the flow $\left(\phi(x_k,t-T_k)\right)_{t\geq T_k}$ for resources, until the moment $T_{k+1}\wedge T$ ;\\
                \quad & \qquad $x_{k+1} \longleftarrow \phi(x_k,T_{k+1}\wedge T-T_k)$ ; \\
                \quad & \qquad \textcolor{blue}{If $T_{k+1} < T$, then} \\
                \quad & \qquad \qquad Simulate $U_{2k}\thicksim\mathcal{U}([0,1])$ ; \\
                \quad & \qquad \qquad \textcolor{blue}{If $U_{2k}\tilde{\alpha}(x_{k+1})\leq \alpha(x_{k+1})$, then} \\
                \quad & \qquad \qquad \qquad $i \longleftarrow 1$ ; \\
                \quad & \qquad \qquad \qquad Simulate $U_{2k+1}\thicksim\mathcal{U}([0,1])$ ;\\
                \quad & \qquad \qquad \qquad $s \longleftarrow \alpha_1(x_{k+1})$ ; \\
                \quad & \qquad \qquad \qquad \textcolor{blue}{While $i<p$ and $U_{2k+1}\alpha(x_{k+1}) > s$, do} \\
                \quad & \qquad \qquad \qquad \qquad $i \longleftarrow i + 1$ ; \\
                \quad & \qquad \qquad \qquad \qquad $s \longleftarrow s + \alpha_i(x_{k+1})$ ; \\
                \quad & \qquad \qquad \qquad \textcolor{blue}{End\_While.} \\
                \quad & \qquad \qquad \qquad $x_{k+1} \longleftarrow x_{k+1} + h_i$ ; \\
                \quad & \qquad \qquad \textcolor{blue}{End\_If.} \\
                \quad & \qquad \textcolor{blue}{End\_If.} \\
                \quad & \qquad $k \longleftarrow k+1$ ; \\
                \quad & \textcolor{blue}{Until $T_k\geq T$.}\\
                \quad & \quad \\
                \hline
            \end{tabular}
        \end{center}
    
\end{document}